\documentclass[preprint,10pt]{elsarticle}

\usepackage{amssymb}
\usepackage[utf8]{inputenc}
\usepackage[english]{babel}
\usepackage{natbib}
\usepackage{hyperref}
\usepackage{amsmath}
\usepackage{longtable}
\usepackage{geometry}

\biboptions{sort&compress}
\journal{Nuclear Physics A}

\begin{document}

\begin{frontmatter}



\title{Study of Decay Modes in Transfermium Isotopes}


\author[a]{U. K. Singh}\author[b]{P. K. Sharma}\author[c]{M. Kaushik}\author[a]{S. K. Jain}\author[d,e]{Dashty T. Akrawy}\author[f]{G. Saxena}
\address[a]{Department of Physics, School of Basic Sciences, Manipal University Jaipur, Jaipur-303007, India}
\address[b]{Govt. Polytechnic College, Rajsamand-313324, India}
\address[c]{S. S. Jain Subodh P. G. College, M. C. A. Institute, Rambagh Circle, Jaipur-302004, India}
\address[d]{Research Center, Salahaddin University, Erbil, Kurdistan, Iraq}
\address[e]{Becquerel Institute for Radiation Research and Measurements, Erbil, Kurdistan, Iraq}
\address[f]{Department of Physics (H\&S), Govt. Women Engineering College, Ajmer-305002, India}

\begin{abstract}
In the unknown territory of transfermium nuclei, the relativistic mean-field (RMF) theory has been applied to probe decay modes which include $\alpha$-decay, spontaneous fission (SF), and a less explored weak-decay. These decay modes are analyzed on equal footing for 101$\leq$Z$\leq$109 and as a consequence, the half-lives for weak-decay are indeed found comparable for several isotopes. Our prediction of decay modes and half-lives are found in excellent agreement with available experimental decay modes and half-lives along with the results of some other theories. Out of $\alpha$, $\beta^+$/EC, $\beta^-$, and SF, the most probable decay mode is anticipated along with its half-life over a wide range of odd and even nuclei to frame a novel sight into terra incognita.
\end{abstract}




\end{frontmatter}


\section{Introduction}
\label{intro}
Extension of the north-east corner of the nuclear chart is one of the prime focuses of nuclear scientists involved in accelerator and detection laboratories.
The knowledge obtained so far in this realm for superheavy nuclei (SHN) is a consequence of coherent progress of experimental and theoretical approaches since last 40 years, which is storied and gathered in Refs~\cite{hofmann2000,hamilton2013,ogan2015,heenen2015,oganrpp2015,oganpt2015,hofmann2016,dull2018,nazar2018,giuliani2019}. The transfermium elements (Z$>$100) up to Z$=$106 were created by irradiations of reactor bred actinide targets with light projectiles, typically O or N \cite{seaborg1985}. Elements up to Z$=$113 are synthesized by cold fusion with Pb or Bi targets and appropriate beams such as Ni and Zn \cite{hofmann2000,morita2004,martens2019} whereas the elements beyond are produced by hot fusion with $^{48}$Ca beams together with suitable actinide targets \cite{ogan2015,heenen2015,oganrpp2015,oganpt2015,giuliani2019,ogan2011,rudolph2013,utyonkov2018,yu2018}.\par

To plan and execute the above-mentioned experiments, a deep knowledge of the decay modes and half-lives of nuclei in a very wide range of nuclear chart is necessary. $\alpha$-decay and spontaneous fission (SF) play a crucial role in the detection of these nuclei in the laboratories as they compete with each other \cite{hofmann2000,hamilton2013,ogan2015,heenen2015,giuliani2019,rudolph2013,utyonkov2018,Oganessian2009}. Another decay mode that is speculated to provide a reach to the nuclei which are not in the original $\alpha$-decay chains is weak-decay ($\beta$-decay) \cite{hofmann2016,karpov2012,zagrebaev2012,ogan2011}. Some theoretical predictions of $\beta$-decay in superheavy nuclei have already been made in Refs. ~\cite{heenen2015,karpov2012,hirsch1993,moller2019,sarriguren2019}. Yet, there is a need of systematic investigation which puts weak-decay on the same ground with $\alpha$-decay and SF so that the chances of weak decay could be explored in superheavy region. With this objective, we employ relativistic mean-field theory (RMF) along with empirical formulas of $\alpha$-decay, $\beta$-decay and SF to calculate probable decay modes and half-lives for the nuclei in the range 101$\leq$Z$\leq$109.\par
\section{Formalism and Calculations}
\subsection{Relativistic Mean-Field Theory} RMF calculations have been carried out using the model Lagrangian density with nonlinear terms both for the ${\sigma}$ and ${\omega}$ mesons as described in detail in Refs.$~$\cite{Singh2013,Yadav2004}.
\begin{small}
\begin{eqnarray}
       {\cal L}& = &{\bar\psi} [\imath \gamma^{\mu}\partial_{\mu}
                  - M]\psi\nonumber\\
                  &&+ \frac{1}{2}\, \partial_{\mu}\sigma\partial^{\mu}\sigma
                - \frac{1}{2}m_{\sigma}^{2}\sigma^2- \frac{1}{3}g_{2}\sigma
                 ^{3} - \frac{1}{4}g_{3}\sigma^{4} -g_{\sigma}
                {\bar\psi}  \sigma  \psi\nonumber\\
               &&-\frac{1}{4}H_{\mu \nu}H^{\mu \nu} + \frac{1}{2}m_{\omega}
                  ^{2}\omega_{\mu}\omega^{\mu} + \frac{1}{4} c_{3}
                 (\omega_{\mu} \omega^{\mu})^{2}
                  - g_{\omega}{\bar\psi} \gamma^{\mu}\psi
                 \omega_{\mu}\nonumber\\
              &&-\frac{1}{4}G_{\mu \nu}^{a}G^{a\mu \nu}
                 + \frac{1}{2}m_{\rho}
                 ^{2}\rho_{\mu}^{a}\rho^{a\mu}
                  - g_{\rho}{\bar\psi} \gamma_{\mu}\tau^{a}\psi
                 \rho^{\mu a}\nonumber\nonumber\\
               &&-\frac{1}{4}F_{\mu \nu}F^{\mu \nu}
                 - e{\bar\psi} \gamma_{\mu} \frac{(1-\tau_{3})}
                 {2} A^{\mu} \psi\,\,,
\end{eqnarray}
\end{small}
where the field tensors $H$, $G$ and $F$ for the vector fields are
defined by
\begin{small}
\begin{eqnarray}
                 H_{\mu \nu} &=& \partial_{\mu} \omega_{\nu} -
                       \partial_{\nu} \omega_{\mu}\nonumber\\
                 G_{\mu \nu}^{a} &=& \partial_{\mu} \rho_{\nu}^{a} -In t
                       \partial_{\nu} \rho_{\mu}^{a}
                     -2 g_{\rho}\,\epsilon^{abc} \rho_{\mu}^{b}
                    \rho_{\nu}^{c} \nonumber\\
                  F_{\mu \nu} &=& \partial_{\mu} A_{\nu} -
                       \partial_{\nu} A_{\mu}\,\,\nonumber\
\end{eqnarray}
\end{small}
and other symbols have their usual meaning. The corresponding
Dirac equations for nucleons and Klein-Gordon equations for mesons
obtained with the mean-field approximation are solved by the
expansion method on the widely used axially deformed
Harmonic-Oscillator basis \cite{Geng2003,Gambhir1989}. The quadrupole
constrained calculations have been performed for all the nuclei
considered here in order to obtain their potential energy surfaces
(PESs) and determine the corresponding ground-state deformations
\cite{Geng2003,Flocard1973}. For nuclei with an odd number of nucleons,
a simple blocking method without breaking the time-reversal
symmetry is adopted \cite{Geng2003wt,Ring1996}.

In the calculations we use for the pairing interaction a delta force, i.e., V = -V$_0 \delta(r)$ with the strength V$_0$ = 350 MeV-fm$^3$ which has been used in Refs.$~$ \cite{Yadav2004,Saxena2017} for the successful description of bubble nuclei \cite{saxena,saxena1,saxenajpg} and also in superheavy nuclei \cite{saxenaijmpe2018,saxenaijmpe2019}. Apart from its simplicity, the applicability and justification of using such a $\delta$-function form of interaction has been discussed in Ref.$~$\cite{Dobaczewski1983}, whereby it has been shown in the context of HFB calculations that the use of a delta force in a finite space simulates the effect of finite range interaction in a phenomenological manner (see also \cite{Bertsch1991} for more details).

Whenever the zero-range $\delta$ force is used either in the BCS
or the Bogoliubov framework, a cutoff procedure must be applied,
i.e. the space of the single-particle states where the pairing
interaction is active must be truncated. This is not only to
simplify the numerical calculation but also to simulate the
finite-range (more precisely, long-range) nature of the pairing
interaction in a phenomenological way \cite{Dobaczewski1995,Goriely2002}. In
the present work, the single-particle states subject to the
pairing interaction is confined to the region satisfying
\begin{small}
\begin{equation}
\epsilon_i-\lambda\le E_\mathrm{cut},
\end{equation}
\end{small}
where $\epsilon_i$ is the single-particle energy, $\lambda$
the Fermi energy, and $E_\mathrm{cut} = 8.0$ MeV. The center-of-mass correction is approximated by
\begin{small}
\begin{equation}
E_{\textrm{cm}} = -\frac{3}{4}41A^{-1/3},
\end{equation}
\end{small}
which is often used in the relativistic mean-field theory among
the many recipes for the center-of-mass correction
\cite{Bender1999}. For further details of these formulations, we refer the
reader to Refs.$~$\cite{Gambhir1989,Singh2013,Geng2003}.

\subsection{$\alpha$-Decay}
The energy release $Q_\alpha$ in ground-state to
ground-state decay is obtained from mass excesses or total binding
energies through
\begin{small}
\begin{eqnarray}
         Q_\alpha(Z, N) & = & M(Z, N) - M(Z-2, N-2) -  M(2, 2) \nonumber\\
& =& B.E.(Z-2, N-2) + B.E.(2, 2) - B.E.(Z, N)
         \label{qalpha}
\end{eqnarray}
\end{small}
where the $^{4}He$ mass excess M(2,2) is 2.42 MeV and the binding
energy B.E.(2,2) is 28.30 MeV. To calculate log$_{10}T_{\alpha}$, we use recently reported modified Royer formula by Akrawy \textit{et al.} \cite{Akrawy2017}.
\begin{small}
 \begin{equation}
 log_{10}T_{\alpha}(sec) = a + bA^{1/6}\sqrt{Z} + \frac{cZ}{\sqrt{Q_{\alpha}}}+ dI + eI^{2}
 \label{alpha}
\end{equation}\end{small}
where I $=$ $\frac{N-Z}{A}$ and the constants a, b, c, d, and e are\\
\begin{table}[!htbp]
\centering
\resizebox{0.6\textwidth}{!}{%
{\begin{tabular}{cccccc}
\hline
 \multicolumn{1}{c}{Nuclei (Z$-$N)}&
 \multicolumn{1}{c}{a}&
 \multicolumn{1}{c}{b}&
 \multicolumn{1}{c}{c}&
 \multicolumn{1}{c}{d}&
 \multicolumn{1}{c}{e}\\
 \hline
 $e-e$&-27.837&-0.9420&1.5343&-5.7004&8.785\\
 $o-e$&-26.801&-1.1078&1.5585&14.8525&-30.523\\
 $e-o$&-28.225&-0.8629&1.5377&-21.145&53.890\\
 $o-o$&-23.635&-0.891&1.404&-12.4255&36.9005\\
\hline
\end{tabular}}}
\end{table}

\subsection{Spontaneous Fission}
The decay mode which is equally important as that of $\alpha$-decay in superheavy region is the spontaneous fission (SF), for which the calculation of half-life is proposed very first by Swiatecki\cite{wjswiatecki1955} based on the fission barrier heights and the values of the fissility parameter $Z^2/A$ and subsequently many other attempts \cite{dwdorn1961,cxu2005} are made to improve the formula. For our investigation, we use the formula given by Karpov \textit{et al.} \cite{karpov2012}
\begin{small}
\begin{eqnarray}
        log_{10}T_{SF}(sec) & = & 1146.44 - 75.3153Z^2/A + 1.63792(Z^2/A)^2 - 0.0119827 (Z^2/A)^3\nonumber\\
& &+B_f  (7.23613 - 0.0947022Z^2/A)\nonumber\\
& &+\begin{cases}
                  \mbox{0, Z and N are even}\\
                  \mbox{1.53897, A is odd}\\
                  \mbox{0.80822, Z and N are odd}.
                  \end{cases}
                           \label{TSF}
\end{eqnarray}\end{small}
Here $B_f$ is the fission barrier, which is calculated as a sum of the liquid-drop barrier $B_f(LDM)$ and the ground state
shell correction $\delta U(g.s.)$, i.e. $B_f$ = $B_f (LDM)$ + $\delta U(g.s.)$ \cite{karpov2012}. For our calculation, we take fission barrier $B_f$ directly from the Ref.$~$ \cite{moller2009}.

\subsection{$\beta$-Decay \& Electron Capture (Weak-decay)}
The energy released in ground-state to ground-state electron
decay ($\beta$-decay) is given in terms of the atomic mass excess $M(Z,N)$ or the
total binding energy $B.E.(Z,N)$ by
\begin{small}
\begin{equation}
        \begin{aligned}[b]
Q_{\beta^-} & =  M(Z, N) - M(Z+1, N-1) \\
& = B.E.(Z+1, N-1) - B.E.(Z, N) + M_n -M _H
        \end{aligned}\label{qbetaminus}
\end{equation}
\end{small}
whereas in positron decay ($\beta^+$-decay) it is
\begin{small}
\begin{eqnarray}
Q_{\beta^+} & = & M(Z, N) - M(Z-1, N+1) - 2m_0c^2 \nonumber \\
& = &B.E.(Z-1, N+1) - B.E.(Z, N) + M _H -M_n -2m_0c^2
         \label{qbetaplus}
\end{eqnarray}
\end{small}
For calculating half-lives for $\beta^+$-decay, electron capture must also be considered because in some cases $\beta^+$ decay is energetically forbidden and electron capture (EC) is possible. The energy released in ground-state to ground-state electron capture (EC) is
\begin{small}
\begin{eqnarray}
Q_{EC} & = & M(Z, N) - M(Z-1, N+1) - \mbox{B.E(electron)} \nonumber \\
& = &B.E.(Z-1, N+1) - B.E.(Z, N) + M_H -M_n - \mbox{B.E(electron)}
         \label{qec}
\end{eqnarray}
\end{small}
so that
\begin{small}
\begin{eqnarray}
         Q_{EC} & =  Q_{\beta^+} + 2m_0c^2 - \mbox{B.E(electron)}
         \end{eqnarray}
\end{small}
To look into the possibility of $\beta$-decay, which is found to be very important for transfermium isotopes ~\cite{heenen2015,karpov2012,hirsch1993,moller2019,sarriguren2019}, we will adopt the empirical formula of Fiset and Nix \cite{Fiset1972} for estimating
the $\beta$-decay half-lives. It is worthy to note that this formula of $\beta$-decay has recently been used in one of our work \cite{saxenaijmpe2019} and the work by Ikram \textit{et al.} \cite{Ikram2017}. However, it should be noted here that nuclear structure that generates the energy
distribution of the GT strength, plays a very important role for $\beta$-decay and hence for an accurate study of $\beta$-decay one has to consider the structure of parent and daughter nuclei ~\cite{heenen2015,sarriguren2019}. Consequently, the calculation of weak-decay rates requires a knowledge of the final states and of the nuclear matrix elements connecting them to the parent ground states. In practice, even after some approximations are made, there is still some task involved and the results are bound to show some model-dependence \cite{heenen2015}. Therefore, to visualize the probability of $\beta$-decay in a more general manner for transfermium isotopes, we follow Fiset and Nix \cite{Fiset1972} from which if $\beta^{\pm}-decay$ or electron capture to the ground state of the daughter nucleus occurs then the inverse half-lives can be written as:
\begin{small}
\begin{eqnarray}
\frac{1}{T_{\beta}}&=&\frac{1}{f_t}{f(Z_d,W_{\beta})}=\frac{1}{f_t}{C(Z_d,W_{\beta})}{f(0, W_{\beta})}\nonumber\\
&\approx&\frac{1}{f_t}\;\frac{1}{30}{C(Z_d,W_{\beta})}(W_{\beta}/m_e)^5,
          \label{eqbeta}
\end{eqnarray}
\noindent
\begin{eqnarray}
       \frac{1}{T_{EC}}&\approx&\frac{1}{f_t} 2\pi(\alpha Z_K)^{2s+1} \left(\frac{2R_0}{\hbar c/ m_e}\right)^{2s-2}\nonumber\\ &&\times\frac{1+s}{\Gamma(2s+1)}\left[\frac{Q_{EC}}{m_e}-(1-s)\right]^2,
            \label{eqEC}
\end{eqnarray}
\end{small}
where the last form of Equation \ref{eqbeta} is valid for $W_{\beta}\gg m_e$. Here in Equations \ref{eqbeta},\ref{eqEC}, $Z_d$ is the proton
number of the daughter nucleus, and $Z_K$ is the effective charge of the parent nucleus for an electron in the K-shell; it is given approximately by $Z_K= Z_P - 0.35$, where $Z_p$ is the proton number of the parent nucleus. The energy $W_{\beta}$ is sum of energy of the emitted $\beta-particle$ and its rest mass $m_e$ i.e.
$W_{\beta} = Q_{\beta}+m_{e}$. Also, the quantity s is given by $s = [ 1 -(\alpha Z_k)^2]^{\frac{1}{2}}$ and represents the rest mass of an electron minus its binding energy in the K-shell, in units of $m_e$. The quantity $\alpha$ is the fine-structure constant, and $R_0$ is the nuclear radius, which is taken to be
$R_0 = 1.2249 A^{\frac{1}{3}} fm$.\par

The function $C(Z_d, W_{\beta}) = f(Z_d, W_{\beta})/f(0, W_{\beta})$ accounts for the increase in the
$\beta-decay$ rate arising from the nuclear Coulomb field. The Fermi integral $f(Z_d, W_{\beta})$ arises from integration over the density of states available to
the emitted $\beta$-particle and neutrino. This function is absent in electron capture, where
the electron is initially in a definite atomic state and consequently the phase-space
volume is determined entirely by the energy of the emitted neutrino. This is responsible
for the difference in the energy dependences of the half-lives in Eqs. \ref{eqbeta},\ref{eqEC}.
$\beta$-decay and electron capture occur not only to the ground states of the daughter
nuclei but also to excited states. Following Seeger \textit{et al.} \cite{Seeger1965}, Eqs. \ref{eqbeta},\ref{eqEC} can be integrated
over the excitation energy E in the daughter nucleus, under the assumption that $1/3$ of
the states in the daughter nucleus are available for such transitions. The energy dependence
of the function $C(Z_d, W_{\beta})$ is neglected. This leads to
\begin{small}
\begin{eqnarray}
 \frac{1}{T_{\beta}}& \approx & \frac{1}{30}\frac{1}{f_t}{C(Z_d)}\frac{1}{m_e^5}\int_{0}^{Q_{\beta}}(W_{\beta}-E)^5\frac{1}{3}\rho dE \nonumber\\ &\approx&\frac{1}{540}\frac{1}{f_t}{C(Z_d)}\frac{\rho}{m_e^5}(W_\beta^6-m_e^6)
 \end{eqnarray}

\begin{eqnarray*}
 \frac{1}{T_{EC}}&\approx&\frac{1}{f_t} 2\pi(\alpha Z_K)^{2s+1} \left(\frac{2R_0}{\hbar c/ m_e}\right)^{2s-2}\frac{1+s}{\Gamma(2s+1)}\frac{1}{m_e^2}\\ &&\times\int_{0}^{Q_{EC.-(1-s)m_e}}\left[Q_{EC}-(1-s)m_e-E\right]^2\frac{1}{3}\rho dE\\
&\approx&\frac{1}{9} \frac{1}{f_t} 2\pi(\alpha Z_K)^{2s+1}\left(\frac{2R_0}{\hbar c/ m_e}\right)^{2s-2}\frac{1+s}{\Gamma(2s+1)}\frac{\rho}{m_e^2}\\
&&\times\left[Q_{EC}-(1-s)m_e\right]^3.
\end{eqnarray*}
\end{small}
For the average density of states $\rho$ in the daughter nucleus, we use the empirical
results given by Seeger \textit{et al.} \cite{Seeger1965}, which are mentioned in Table 1.

\begin{table}[!htbp]
 \centering
 \caption{Density of nuclear states ($e^{-A/290}\times$ number of states within 1 MeV of ground state). }
 \resizebox{0.5\textwidth}{!}{%
 \begin{tabular}{c c c c c}

 \hline
 Nuclear & \multicolumn{3}{c}{Spherical} & Deformed \\
 \cline{2-4}

 species & doubly & singly & neither  &  \\
 &  magic &  magic &  magic &  \\
 \hline
 even & 0.22 & 0.97 & 1.36 & 2.73 \\
 odd-mass & 0.60 & 1.67 & 5.0 & 8.6 \\
 odd & 7.5 & 8.6 & 15.0 & 15.0 \\

 \hline

\end{tabular}}
\end{table}

Upon inserting the values $f_t = 10^{6.5} s$ and $C(Z_d) = 10^{1.5}$, we are led finally to
\begin{small}
\begin{equation}
        \begin{aligned}[b]
T_{\beta}&=  \frac{540 m_e^5}{\rho(W_\beta^6-m_e^6)}\times 10^{5.0}s
           \end{aligned} \label{tbeta}
\end{equation}
\begin{eqnarray}
 T_{EC} &= &\frac{9 m_e^2}{2\pi(\alpha Z_K)^{2s+1}\rho\left[Q_{EC}-(1-s)m_e\right]^3} \left(\frac{2R_0}{\hbar c/ m_e}\right)^{2-2s} \nonumber\\
 &&\times\frac{\Gamma(2s+1)}{1+s}\times10^{6.5}s.
             \label{tecfinal}
\end{eqnarray}
\end{small}
We have concentrated here on electron capture from the K-shell because it is usually the
predominant $\beta$-decay process, however, electron capture from the $L_1$ shell also occurs for superheavy nuclei with a relative
probability of about 20\%. In this paper, we will follow Eqn. \ref{tbeta} to calculate half-lives for $\beta^-$-decay and $\beta^+$-decay whereas the Eqn. \ref{tecfinal} will be used to calculate half-life for electron capture. The half-life with respect to $\beta^+$/EC-decay is given by
\begin{small}
\begin{eqnarray}
\frac{1}{T_{\beta^+/EC}} =  \frac{1}{T_{\beta^+}} + \frac{1}{T_{EC}}
 \label{tbetaplusec}
\end{eqnarray}
\end{small}
\section{Results and discussions}
The term "transfermium" describes the elements with Z$>$100(Fermium). Therefore, in this paper we have considered the isotopes of Md, No, Lr, Rf, Db, Sg, Bh,
Hs, and Mt (101$\leq$Z$\leq$109) that can be produced in the frontier of cold and hot fusion reactions and recently attracted the superheavy world with a great interest \cite{yu2018,martens2019,hong2016,hong2017}. Though the superheavy world has reached the Z$=$118 \cite{ogan2006}, still there are a lot of nuclei that are yet to be explored through their decay modes, lifetimes, and other properties. Therefore, this article aims to probe the undiscovered land of elements of 101$\leq$Z$\leq$109 including odd and even nuclei in the range of $^{235-268}$Md, $^{238-268}$No, $^{241-270}$Lr, $^{243-272}$Rf, $^{245-272}$Db, $^{248-276}$Sg, $^{250-278}$Bh, $^{253-282}$Hs, and $^{255-284}$Mt. For all these nuclei, the calculations are done using two parameters i.e. NL3* \cite{nl3star} and TMA \cite{sugaTMA} of relativistic mean-field theory (RMF) \cite{Yadav2004,Saxena2017,saxenaijmpe2018,saxenaijmpe2019} as explained above. At places, we will compare our results with available experimental data \cite{nndc}. In addition, few of our results are also compared with Hartree-Fock-Bogoliubov (HFB) mass model with HFB-24 functional \cite{hfbxu}, relativistic continuum Hartree-Bogoliubov (RCHB) theory with the relativistic density functional PC-PK1 \cite{rchb2018}, nuclear mass table with the global mass formula WS4 \cite{ws42014}, and, recently reported Finite Range Droplet Model (FRDM) calculations \cite{moller2019}.\par
\begin{figure}[ht]
\centering
\includegraphics[width=0.6\textwidth]{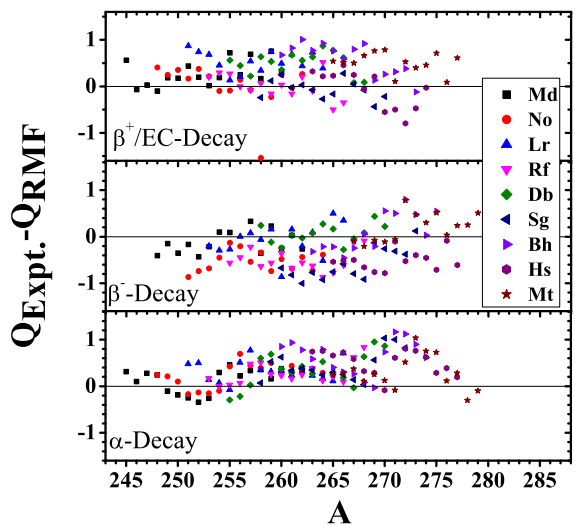}
\caption{(Colour online) Difference of calculated and experimental \cite{nndc} Q-values for $\alpha$, $\beta^-$ and $\beta^+$/EC-decays.} \label{fig1}
\end{figure}
\begin{table*}[!htbp]
\caption{Root mean square error (RMSE) of Q values for $\alpha$, $\beta^-$ and $\beta^+$/EC-decays for each isotopic chain.}
\centering
\def\arraystretch{1.0}
\resizebox{1.0\textwidth}{!}{%
{\begin{tabular}{cccccccccccccccccc}
 \hline
 \multicolumn{1}{c}{Nucleus}&
 \multicolumn{5}{c}{Q$_{\alpha}$}&
  \multicolumn{1}{c}{}&
 \multicolumn{5}{c}{Q$_{\beta^{-}}$}&
   \multicolumn{1}{c}{}&
 \multicolumn{5}{c}{Q$_{\beta^{+}/EC}$}\\
 \cline{2-6} \cline{8-12}  \cline{14-18}
&TMA&NL3*&RCHB&WS4&FRDM&&TMA&NL3*&HFB&RCHB&WS4&&TMA&NL3*&RCHB&WS4&FRDM\\
 \cline{1-6} \cline{8-12}  \cline{14-18}
 Md   &  0.29  &    0.57& 1.24 & 0.24  & 0.28  &&  0.25  &  0.36  &  0.24& 0.83&0.19 &&   0.39  &  0.50 & 1.10&0.40  &  0.13  \\
 No   &  0.35  &    0.49& 0.86 & 0.25  & 0.32  &&  0.57  &  0.53  &  0.26& 1.04&0.34 &&   0.25  &  0.29 & 0.84&0.18  &  0.18   \\
 Lr   &  0.27  &    0.41& 0.77 & 0.27  & 0.48  &&  0.26  &  0.46  &  0.26& 0.74&0.24 &&   0.57  &  0.63 & 1.04&0.34  &  0.11   \\
 Rf   &  0.36  &    0.43& 0.86 & 0.22  & 0.46  &&  0.55  &  0.56  &  0.36& 1.06&0.28 &&   0.26  &  0.36 & 0.74&0.24  &  0.26     \\
 Db   &  0.49  &    0.55& 1.05 & 0.25  & 0.50  &&  0.21  &  0.81  &  0.54 & 0.59&0.14 &&   0.56  &  0.46 & 1.06&0.28  &  0.13      \\
 Sg   &  0.54  &    0.88& 1.33 & 0.28  & 0.47  &&  0.67  &  0.97  &  0.38& 1.12&0.46 &&   0.21  &  0.73 &0.59&0.14  &  0.23\\
 Bh   &  0.72  &    0.53& 1.69 & 0.20  & 0.44  &&  0.41  &  0.79  &  0.50& 0.56&0.21 &&   0.67  &  0.97 &1.12&0.46  &  0.28  \\
 Hs   &  0.58  &    0.54& 1.90 & 0.16  & 0.39  &&  0.41  &  0.54  &  0.44& 0.65&0.36 &&   0.55  &  0.79 &0.56&0.20  &  0.32    \\
 Mt   &  0.45  &    0.80& 2.49 & 0.32  & 0.60  &&  0.36  &  0.98  &  0.43 & 0.58&0.23 &&   0.55  &  0.63 &0.65&0.36  &  0.23      \\
 \hline
\end{tabular}}
}
\end{table*}
First, to validate our results of RMF, we compute the difference of our calculated Q-values using the TMA parameter with the available experimental data \cite{nndc}. The Q-values are calculated for $\alpha$, $\beta^-$, $\beta^+$, and EC-decays using Eqns. \ref{qalpha},\ref{qbetaminus},\ref{qbetaplus},\ref{qec} and the differences are plotted in Fig. \ref{fig1}. In this figure, out of $\beta^+$ and EC-decay, we plot Q-values for EC-decay only (mentioning $\beta^+$/EC decay). From Fig. \ref{fig1} it is indulging to note that our results of Q-values are in excellent agreement as for most of the nuclei the differences of Q-values i.e. Q$_{Expt.}$$-$Q$_{Calc.}$$<$1.0 MeV, which manifests our predictions for decay modes. At this point, it will be more appropriate to compare our calculations with other parameters/theories. Therefore, in Table 2, we have shown the root mean square error (RMSE) of Q values (Q$_{Expt.}$$-$Q$_{RMF}$) for $\alpha$, $\beta^-$ and $\beta^+$/EC-decays for each isotopic chain. The comparison is done with another parameter of RMF i.e. NL3* also with other theories viz. HFB \cite{hfbxu}, RCHB(PC-PK1) \cite{rchb2018}, WS4 \cite{ws42014} and  FRDM \cite{moller2019}. As per the table, it is gratifying to note that RMSE from both the parameters of RMF is found reasonably lower similar to the other theories. However, it is also important to point out here that the TMA parameter, which is a mass-dependent parameter, is found more appropriate in this region as compared to the non-linear variant of the parameter (NL3*). Therefore, in the following, TMA parameter will be considered for the more accurate prediction of decay modes.\par

As already mentioned in the introduction, the competition between the different decay modes is important to determine
the stability of a particular nucleus and consequently to provide a reach from a given hot-fusion reaction \cite{hamilton2013,ogan2015,heenen2015,giuliani2019}.
Therefore, we employ the calculated Q-values from our theory and the formula mentioned in section II to calculate half-lives of $\alpha$, $\beta^\pm$, and EC-decays using Eqns. \ref{alpha}, \ref{tbeta} and \ref{tecfinal}, respectively. Similarly, the half-life for spontaneous fission is calculated using Eqn. \ref{TSF}. These all half-lives are tabulated together to demonstrate their competence and to bring the most favourable or probable decay mode for each nucleus in considered isotopic chains. In order to compare and validate our results of decay modes and half-life of probable decay mode, we first compare our theoretical prediction with available experimental data \cite{nndc} in Table 3. For this comparison only those nuclei are taken into consideration from the experimental database \cite{nndc} which have more than 50$\%$ probability of any particular decay mode or the decay modes are comparable.\par
\begin{center}
\small
\setlength{\tabcolsep}{3pt}
\begin{longtable}{|c|c|c|c|c|c|c|c|c|c|}
\caption{Comparison of decay-modes and half-lives with our theoretical prediction for transfermium isotopes. Experimental decay modes and half-lives are taken from \cite{nndc}.}\\
\hline
 \multicolumn{1}{|c}{Nucleus}&
 \multicolumn{5}{|c}{Log T$_{1/2}$}&
 \multicolumn{1}{|c}{Predicted}&
 \multicolumn{1}{|c}{Expt.}&
 \multicolumn{1}{|c}{Predicted}&
  \multicolumn{1}{|c|}{Expt.}\\
 \cline{2-6}
 \multicolumn{1}{|c}{}&
 \multicolumn{1}{|c}{Log T$_\alpha$}&
  \multicolumn{1}{|c}{Log T$_{\beta^-}$}&
   \multicolumn{1}{|c}{Log T$_{\beta^+}$}&
  \multicolumn{1}{|c}{Log T$_{EC}$}&
   \multicolumn{1}{|c}{Log T$_{SF}$}&
 \multicolumn{1}{|c}{Decay-Mode}&
 \multicolumn{1}{|c}{Decay-Mode}&
 \multicolumn{1}{|c}{T$_{1/2}$}&
  \multicolumn{1}{|c|}{T$_{1/2}$}\\
\hline
\endfirsthead
\multicolumn{10}{c}%
{\tablename\ \thetable\ -- \textit{Continued from previous page}} \\
\hline
 \multicolumn{1}{|c}{Nucleus}&
 \multicolumn{5}{|c}{Log T$_{1/2}$}&
 \multicolumn{1}{|c}{Predicted}&
 \multicolumn{1}{|c}{Expt.}&
 \multicolumn{1}{|c}{Predicted}&
  \multicolumn{1}{|c|}{Expt.}\\
 \cline{2-6}
 \multicolumn{1}{|c}{}&
 \multicolumn{1}{|c}{Log T$_\alpha$}&
  \multicolumn{1}{|c}{Log T$_{\beta^-}$}&
   \multicolumn{1}{|c}{Log T$_{\beta^+}$}&
  \multicolumn{1}{|c}{Log T$_{EC}$}&
   \multicolumn{1}{|c}{Log T$_{SF}$}&
 \multicolumn{1}{|c}{Decay-Mode}&
 \multicolumn{1}{|c}{Decay-Mode}&
 \multicolumn{1}{|c}{T$_{1/2}$}&
  \multicolumn{1}{|c|}{T$_{1/2}$}\\
\hline
\endhead
\hline \multicolumn{10}{r}{\textit{Continued on next page}} \\
\endfoot
\hline
\endlastfoot
$^{245}$Md &   0.48&	-&   1.71&	2.49 &	3.51  &   $\alpha$                   & SF/$\alpha$  &  3.02s  & (0.90$\pm$0.25)ms\\
$^{246}$Md &   0.73&	-&	0.68&	1.88 &	4.27  & $\beta^{+}/EC$/$\alpha$      &   $\alpha$  &  4.73s     &   (0.9$\pm$0.2)s \\
$^{247}$Md &   1.03&	-&	1.92&	2.58 & 6.21   &  $\alpha$                   &   $\alpha$  & 10.80s    &   (1.2$\pm$0.1)s      \\
$^{248}$Md &   1.65&	-&	0.99&	2.03 & 6.76   &   $\beta^{+}/EC$/$\alpha$    &   $\alpha$/$\beta^{+}/EC$  & 9.75s    &   $(13^{+15}_{-4})$s   \\
$^{249}$Md &   0.84&	-&	2.47&	2.83 &  8.53   & $\alpha$                   &   $\alpha$/$\beta^{+}/EC$  & 6.89s     &   (21.7$\pm$2.0)s    \\
$^{250}$Md  & 1.51    &  - &   1.54   & 2.28      &   9.68    &    $\alpha$/$\beta^{+}/EC$    &$\beta^{+}/EC$  &  32.28s &$(25^{+10}_{-5})$s   \\
$^{251}$Md  & 1.90    &  - &   3.45   & 3.27      &   11.71   &    $\alpha$                   &$\beta^{+}/EC$  &  79.89s & (4.27$\pm$0.26)m  \\
$^{252}$Md  & 2.57    &  - &   2.24   & 2.60      &   12.20   &    $\beta^{+}/EC$/$\alpha$    &$\beta^{+}/EC$  &  2.89m  & (2.3$\pm$0.8)m  \\
$^{253}$Md  & 3.20    &  - &   4.65   & 3.76      &   12.55   &    $\alpha$/$\beta^{+}/EC$    &$\beta^{+}/EC$  &  26.40m & $(6^{+12}_{-3})$m  \\
$^{254}$Md  & 4.60    &  - &   3.49   & 3.15      &   10.82   &    $\beta^{+}/EC$/$\alpha$    &$\beta^{+}/EC$  &  23.54m & (28$\pm$8)m  \\
$^{255}$Md  & 4.61    &  - &     -    & 6.83      &   10.27   &    $\alpha$                   &$\beta^{+}/EC$  &  11.30h & (27$\pm$2)m  \\
$^{256}$Md  & 4.53    &  - &   4.60   & 3.60      &   8.40    &    $\beta^{+}/EC$             &$\beta^{+}/EC$  &  65.70m &  (77$\pm$2)m \\
$^{257}$Md  & 5.48    &  - &     -    &  -        &   7.42    &    $\alpha$                   &$\beta^{+}/EC$  &  84.79h &  (5.52$\pm$0.05)h \\
$^{258}$Md &   6.74& 4.92&	6.79&	4.27 &  6.19   &  $\beta^{+}/EC$/$\beta^{-}$              &$\beta^{+}/EC$ & 5.17m    &   (57.5$\pm$0.9)m    \\
$^{259}$Md &   6.57&	-&     -&	-    &	6.70   &  $\alpha$/SF               &   SF &        2.50h     &   (1.6$\pm$0.6)h      \\
$^{260}$Md &   8.33 &3.78 &  -&     -&     7.24  &   $\beta^{-}$            &   SF/$\alpha$/$\beta$              &0.07d &(31.8$\pm$0.5)d\\
\hline
$^{250}$No &  0.13&	-&	0.18 & 	0.41 &    4.38  &     $\alpha$/$\beta^{+}/EC$           &        SF           &     1.36s  &        $(4.2^{+1.2}_{-0.9})$$\mu$s \\
$^{251}$No &  0.32&	-&	0.43	&0.57 &    7.61  &    $\alpha$/$\beta^{+}/EC$              &     $\alpha$        &  2.09s     &        (0.80$\pm$0.01)s   \\
$^{252}$No &  0.59&	-&	-0.04&	0.28&     7.22 &      $\beta^{+}/EC$/$\alpha$    &     $\alpha$        &    0.90s   &        (2.44$\pm$0.04)s \\
$^{253}$No &  1.37&	-&	0.29	&0.47 &    10.52 &    $\beta^{+}/EC$            &      $\alpha$/$\beta^{+}/EC$        &  1.95s     &        (1.62$\pm$0.15)m\\
$^{254}$No &  1.68&	-&	-0.46&	0.14&     8.75 &      $\beta^{+}/EC$            &       $\alpha$/$\beta^{+}/EC$       &    0.35s   &        (51$\pm$10)s\\
$^{255}$No  &  3.20    &  - &   0.01  &  0.31   &  9.34    &   $\beta^{+}/EC$ & $\beta^{+}/EC$/$\alpha$  &    1.03s   & (3.52$\pm$0.21)m   \\
$^{256}$No &  3.11&	-&	-	   &-0.64&     6.65 &     $\beta^{+}/EC$            &       $\alpha$      &   0.23s    &        (2.91$\pm$0.05)s\\
$^{257}$No &  3.07&	-&	-0.51&	0.12&     7.10 &      $\beta^{+}/EC$/$\alpha$    &        $\beta^{+}/EC$/$\alpha$    &    0.31s   &        (24.5$\pm$0.5)s\\
$^{258}$No &  3.54&	-&	-    &	-   &     4.06 &       $\alpha$                  &          SF         &    0.96h   &        (1.2$\pm$0.2)ms \\
$^{259}$No &  5.55&	-&	-	   &-  &     6.01 &       $\alpha$                    &         $\alpha$    &   97.60h   &        (58$\pm$5)m\\
$^{260}$No &  4.81&	-&	-	   & -   &     4.57 &      SF                         &         SF          &   10.34h   &        (106$\pm$8)ms\\
$^{262}$No &  6.78&	-&	-	   &  -  &      5.91&      SF                         &          SF         &   9.32d    &        5ms\\
\hline
 $^{252}$Lr  &  1.73 & -    &1.11  & 2.01  &  4.81  &  $\beta^{+}/EC$/$\alpha$   &   $\alpha$  &  12.87s     &$(0.36^{+0.11}_{-0.07})$s\\
 $^{253}$Lr  &  0.80 & -    &2.45  & 2.76  &  6.80  &  $\alpha$                  &   $\alpha$  &  6.24s     &$(0.57^{+0.07}_{-0.06})$s\\
 $^{254}$Lr  &  1.48 & -    &1.40  & 2.15  &  7.77  &  $\beta^{+}/EC$/$\alpha$   &   $\alpha$/$\beta^{+}/EC$  &  25.04s     &(18.4$\pm$1.8)s\\
 $^{255}$Lr  &  1.24 & -    &2.96  & 2.98  &  8.48  &  $\alpha$                  &   $\alpha$  &  17.45s     &(31.1$\pm$1.3)s\\
 $^{256}$Lr  &  2.75 & -    &2.07  & 2.45  &  6.97  &  $\beta^{+}/EC$/$\alpha$   &   $\alpha$/$\beta^{+}/EC$  &  116.26s    &(27$\pm$3)s\\
 $^{257}$Lr  &  2.32 & -    &4.52  & 3.64  &  6.77  &  $\alpha$                  &   $\alpha$  &  208.02s      &4s\\
 $^{258}$Lr  &  1.99 & -    &2.76  & 2.76  &  5.08  &  $\alpha$                  &   $\alpha$  &  97.42s    &(4.1$\pm$0.3)s\\
 $^{259}$Lr  &  2.41 & -    &8.02  & 4.54  &  5.08  &  $\alpha$                  &   $\alpha$  &  256.07s   &(6.2$\pm$0.3)s\\
  $^{260}$Lr &  3.46 & -    &3.77  & 3.19  &  5.27  &  $\alpha$                  &   $\alpha$  &  47.60m  &(180$\pm$30)s\\
  $^{261}$Lr &  3.65 & -    & -    & 6.20  &  6.25  &  $\alpha$                  &   SF        &  4.63m     &(39$\pm$12)m\\
  $^{262}$Lr  &   5.02   &   -    &  4.97  &     3.67    &   6.35    &    $\beta^{+}/EC$ & $\alpha$/$\beta^{+}/EC$   &   1.30h&(4.0h)         \\
$^{266}$Lr &  7.58 & 7.00 & -    & 5.13  &  9.49  &  $\beta^{+}/EC$            &   SF        &  37.86h     &$(11^{+21}_{-5})$h\\
\hline
$^{253}$Rf  &   0.23   & - &   1.55   &  2.32  & 3.09  &  $\alpha$                &   SF/$\alpha$&  1.71s   &  $(48^{+17}_{-10})$$\mu$s\\
$^{254}$Rf  &   -0.17  & - &   3.46   &  3.45  & 2.89  &  $\alpha$                &   SF       &  0.68s   &  (23$\pm$3)$\mu$s\\	
$^{255}$Rf  &   0.78   & - &   2.01   &  2.53  & 5.89  &  $\alpha$                &   SF/$\alpha$&  6.01s   &  (1.68$\pm$0.09)s\\	
$^{256}$Rf  &   0.81   & - &   4.07   &  3.72  & 4.69  &  $\alpha$                &   SF       &  6.45s   &  (6.67$\pm$0.10)ms\\	
$^{257}$Rf  &   2.04   & - &   2.69   &  2.84  & 5.72  &  $\alpha$                &   $\alpha$ &  108.83s  &  $(4.4^{+0.6}_{-0.5})$s\\
$^{258}$Rf  &   1.31   & - &   5.88   &  4.45  & 3.26  &  $\alpha$                &   SF/$\alpha$&  20.29s  &  $(14.7^{+1.2}_{-0.1})$ms\\
$^{259}$Rf  &   1.16   & - &   3.34   &  3.13  & 4.55  &  $\alpha$                &   $\alpha$ &  14.50s  &  (2.4$\pm$0.4)s\\	
$^{260}$Rf  &   1.30   & - &    -     &  5.35  & 2.86  &  $\alpha$                &   SF       &  19.79s  &  (21$\pm$1)ms\\	
$^{261}$Rf  &   2.35   & - &   4.45   &  3.59  & 5.35  &  $\alpha$                &  $\alpha$  &  224.81s &  (68$\pm$3)s\\	
$^{262}$Rf  &   2.63   & - &    -     &   -    & 4.20  &  $\alpha$                &   SF       &  428.21s &  (2.3$\pm$0.4)s\\	
$^{263}$Rf  &   4.45   & - &   7.00   &  4.43  & 6.10  &  $\beta^{+}/EC$/$\alpha$ &   SF       &   7.39h  &  (10$\pm$2)m\\	
$^{265}$Rf  &   5.45   & - &    -     &  5.79  & 8.94  &  $\alpha$                &   SF       &   3.29d  &  $(1^{+12}_{-3})$m\\
\hline
$^{255}$Db  &   -1.27  & -   & 1.58  &  2.29 & 2.83  &   $\alpha$                 &  $\alpha$  &  0.05s   &   $(1.6^{+0.6}_{-0.4})$s\\
$^{256}$Db  &   -0.01  & -   & 0.74  &  1.77 & 3.54  &   $\alpha$                 &  $\alpha$  &  0.99s   &  $(1.6^{+0.5}_{-0.4})$s\\
$^{257}$Db  &   0.27   & -   & 2.00  &  2.48 & 5.06  &   $\alpha$                 &  $\alpha$  &  1.87s   &   (2.3$\pm$0.2)s\\	
$^{258}$Db  &   1.72   & -   & 1.27  &  2.01 & 3.92  &   $\beta^{+}/EC$/$\alpha$  &  $\alpha$  &  18.45s  &   $(4.2^{+0.4}_{-0.3})$s\\
$^{259}$Db  &   1.00   & -   & 2.90  &  2.89 & 4.10  &   $\alpha$                 &  $\alpha$  &  10.05s  &   (0.51$\pm$0.16)s\\	
$^{260}$Db  &   1.07   & -   & 1.72  &  2.22 & 3.79  &   $\alpha$                 &  $\alpha$  &  11.64s  &   (1.52$\pm$0.13)s\\	
$^{261}$Db  &   1.01   & -   & 3.80  &  3.28 & 4.62  &   $\alpha$                 &  $\alpha$  &  10.22s  &   (1.8$\pm$0.4)s\\	
$^{262}$Db  &   2.45   & -   & 2.39  &  2.53 & 4.74  &   $\beta^{+}/EC$/$\alpha$  &  $\alpha$  &  243.38s &   35s\\		
$^{263}$Db  &   2.14   & -   & 4.83  &  3.70 & 5.62  &   $\alpha$                 &  SF/$\alpha$&  139.57s &   $(27^{+10}_{-7})$s\\
$^{267}$Db  &   4.28   & -   &  -    &  5.67 & 10.38 &   $\alpha$                 &  SF        &  315.43m &   $(73^{+350}_{-33})$m\\
$^{268}$Db  &   5.46   & -   & 5.09  &  3.65 & 9.99  &   $\beta^{+}/EC$/$\alpha$  &  SF        &    1.24h &   $(32^{+11}_{-7})$h\\
$^{270}$Db  &   6.43   & 4.88&  -    &  4.35 & 7.70  &   $\beta^{+}/EC$  &  $\alpha$  &    6.28h &   $(1^{+19}_{-4})$h\\

\hline
$^{258}$Sg  &  -0.36 & - &2.87   &3.11  &  1.66   &  $\alpha$  &  SF         &   0.43s    &$(2.9^{+1.3}_{-0.7})$ms\\
$^{259}$Sg  &  0.91  & - &1.79   &2.35  &  3.09   &  $\alpha$  & $\alpha$    &   8.08s    &(0.29$\pm$0.55)s\\	
$^{260}$Sg  &  0.30  & - &3.88   &3.56  &  1.85   &  $\alpha$  &  $\alpha$/SF&   2.02s    &(4.95$\pm$0.33)ms\\	
$^{261}$Sg  &  0.47  & - &2.29   &2.58  &  3.76   &  $\alpha$  &  $\alpha$   &   2.94s    &(178$\pm$14)ms\\	
$^{262}$Sg  &  0.45  & - &4.66   &3.89  &  2.56   &  $\alpha$  &  SF         &   2.81s    &$(6.9^{+3.8}_{-1.8})$ms\\
$^{263}$Sg  &  1.38  & - &2.80   &2.81  &  5.17   &  $\alpha$  & $\alpha$    &   23.85s   &(1.0$\pm$0.2)s\\
$^{264}$Sg	&  1.12	&-	 &5.45	 &4.21	&   3.28  &  $\alpha$  & SF/$\alpha$ &   13.16s  & $(37^{+27}_{-11})$ms\\
$^{265}$Sg  &  2.27  & - &3.59   &3.16  &  6.01   &  $\alpha$  & $\alpha$    &   186.07s  &$(14.4^{+3.7}_{-2.5})$s\\
$^{266}$Sg  &  2.39  & - & -     &5.81  &  6.09   &  $\beta^{+}/EC$/$\alpha$   &  SF         &   245.26s  &$(21^{+20}_{-12})$s\\
$^{269}$Sg  &  4.46  & - & 7.80  &4.44  &  10.47  &  $\alpha$  & $\alpha$    &   7.93h    &$(3.1^{+3.7}_{-1.1})$m\\ 
$^{271}$Sg  &  5.16  & - & -     &6.12  &  8.35   &  $\alpha$  &  $\alpha$/SF&   39.93h   &(2.4$\pm$4.3)m\\	
 \hline
$^{260}$Bh  &  0.65  &  - &  0.61 &1.63  & 1.62   &  $\beta^{+}/EC$/$\alpha$ &  $\alpha$ &     4.07s   &  $(35^{+19}_{-9})$ms\\
$^{261}$Bh  &  -0.16 &  - &  1.86 &2.35  & 3.01   &  $\alpha$                &  $\alpha$ &     0.68s   &  $(11.8^{+3.9}_{-2.4})$ms\\
$^{262}$Bh  &  0.65  &  - &  1.08 &1.85  & 3.10   &  $\alpha$                &  $\alpha$ &     4.48s   &   $(22\pm4)$ms\\	
$^{263}$Bh  &  -0.05 &  - &  2.46 &2.62  & 4.06   &  $\alpha$                &  $\alpha$ &     0.90s   &     \\
$^{264}$Bh  &  1.59  &  - &  1.58 &2.09  & 4.16   &  $\beta^{+}/EC$/$\alpha$ &  $\alpha$ &     37.65s  &   $(0.44^{+0.60}_{-0.16})$s\\
$^{265}$Bh  &  0.63  &  - &  3.13 &2.92  & 4.98   &  $\alpha$                &  $\alpha$ &    4.23s    &   $(0.9^{+0.7}_{-0.3})$s\\
$^{266}$Bh  &  1.70  &  - &  1.90 &2.24  & 5.66   &  $\alpha$                &  $\alpha$ &    50.37s   &   $(1.7^{+8.2}_{-0.8})$s\\
$^{267}$Bh  &  1.56  &  - &  3.97 &3.28  & 7.47   &  $\alpha$                &  $\alpha$ &    36.65s   &   $(17^{+14}_{-6})$s\\
$^{268}$Bh  &  2.26  &  - &  2.64 &2.57  & 7.73   &  $\alpha$                &  $\alpha$ &    183.18s  &    \\
$^{269}$Bh  &  2.60  &  - &  5.19 &3.77  & 10.01  &  $\alpha$                &  $\alpha$ &    398.86s  &      \\
$^{270}$Bh  &  4.01  &  - &  3.38 &2.90  & 9.85   &  $\beta^{+}/EC$          &  $\alpha$ &    786.19s  &   $(60^{+29}_{-3})$s\\
$^{271}$Bh  &  3.86  &  - &  -    &4.66  & 9.89   &  $\alpha$                &  $\alpha$ &     2.02h   &  	   \\
$^{272}$Bh  &  4.46  &  - &  4.40 &3.32  & 7.97   &  $\beta^{+}/EC$/$\alpha$ &  $\alpha$ &    34.67m   &  $(10^{+12}_{-4})$s\\
$^{274}$Bh  &  3.95  &  - &  5.00 &3.55  & 5.22   & $\beta^{+}/EC$/$\alpha$  &  $\alpha$ &    59.43m   &  $(0.9^{+4.2}_{-0.4})$m\\
\hline
$^{263}$Hs  &  -0.43  &  - &1.50  &2.14  & 2.37  &  $\alpha$ & $\alpha$ & 0.37s   &  $(0.74^{+0.48}_{-0.21})$ms\\
$^{264}$Hs  &  -0.54  &  - &3.17  &3.18  & 1.42  &  $\alpha$ & $\alpha$/SF & 0.29s   &  $(0.8^{+3.9}_{-2.4})$ms\\
$^{265}$Hs  &  0.00   &  - &1.90  &2.33  & 3.58  &  $\alpha$ & $\alpha$ & 0.99s   &  (1.9$\pm$0.2)ms\\	
$^{266}$Hs  &  0.00   &  - &3.95  &3.52  & 2.27  &  $\alpha$ & $\alpha$ & 1.00s   &  $(2.3^{+1.3}_{-0.6})$ms\\
$^{267}$Hs  &  0.95   &  - &2.36  &2.54  & 4.68  &  $\alpha$ & $\alpha$ & 8.96s   &  $(52^{+13}_{-8})$ms\\
$^{268}$Hs  &  0.87   &  - &4.71  &3.84  & 3.75  &  $\alpha$ & $\alpha$ & 7.36s   &  $(0.4^{+1.8}_{-0.2})$s\\
$^{269}$Hs  &  2.11   &  - &3.08  &2.87  & 6.94  &  $\alpha$ & $\alpha$ & 128.43s &  $(9.7^{+9.3}_{-0.3})$s\\
$^{270}$Hs  &  1.25   &  - &6.01  &4.36  & 6.69  &  $\alpha$ & $\alpha$ & 17.72s  &  22s		\\
$^{273}$Hs  &  2.41   &  - &4.73  &3.56  & 7.59  &  $\alpha$ & $\alpha$ & 258.14s &  $(0.76^{+0.71}_{-0.24})$s\\
$^{275}$Hs  &  1.74   &  - &5.53  &3.87  & 5.37  &  $\alpha$ & $\alpha$ & 55.03s  & $(0.15^{+0.27}_{-0.06})$s\\
$^{277}$Hs  &  2.64   &  - & -    &4.53  & 4.58  &  $\alpha$ & $\alpha$ & 431.98s &  $(3^{+15}_{-1})$ms\\
\hline
$^{266}$Mt &  -1.52  &  - &0.53 &  1.52 &  0.84 &  $\alpha$ & $\alpha$ &  0.03s    &  $(1.7^{+1.8}_{-1.6})$ms\\
$^{267}$Mt &  -2.10  &  - &1.67 &  2.19 &  2.26 &  $\alpha$ & $\alpha$ &  0.01s    &  		      \\
$^{268}$Mt &  -1.21  &  - &0.84 &  1.67 &  2.48 &  $\alpha$ & $\alpha$ &  0.06s    &  $(21^{+8}_{-5})$ms \\
$^{270}$Mt &  -0.05  &  - &1.29 &  1.88 &  5.56 &  $\alpha$ & $\alpha$ &  0.89s    &  $(5^{+24}_{-3})$ms\\
$^{274}$Mt &  0.43   &  - &2.38 &  2.38 &  6.43 &  $\alpha$ & $\alpha$ &  2.68s    &  $(0.44^{+0.81}_{-0.17})$s\\
$^{275}$Mt &  -0.14  &  - &5.41 &  3.79 &  5.77 &  $\alpha$ & $\alpha$ &  0.73s    &  $(9.7^{+46.0}_{-0.4})$ms\\
$^{276}$Mt &  0.07   &  - &2.81 &  2.58 &  3.92 &  $\alpha$ & $\alpha$ &  1.18s    &  $(0.72^{+0.87}_{-0.25})$s\\
$^{277}$Mt &  0.20   &  - & -   &  4.60 &  3.20 &  $\alpha$ & SF &        1.59s    &  $(5^{+9}_{-2})$s\\
$^{278}$Mt &  0.19   &  - &3.12 &  2.71 &  2.09 &  $\alpha$ & $\alpha$ &  1.56s    &  $(8^{+37}_{-4})$s\\
\hline
\end{longtable}
\end{center}
\normalsize

In Table 3, half-lives of only those nuclei are mentioned in which decay mode is energetically possible (Q$>$0). Even though $\beta^{+}$-decay and EC-decay are very difficult to distinguish, the half-lives for both these decays are tabulated separately considering the fact that T$_{\beta^+}$ and  T$_{EC}$ are calculated using separate formula \cite{Fiset1972} as mentioned in Eqns. \ref{tbeta} and \ref{tecfinal}. However, in the predicted decay mode $\beta^{+}/EC$ is mentioned always even if any of the two decay modes is possible. From Table 3, a purposeful comparison for T$_\alpha$, T$_{\beta^-}$, T$_{\beta^+}$, T$_{EC}$
and T$_{SF}$ can be done easily and the lowest value of half-life can be determined to lead to the most favourable decay mode, tabulated along with the comparison with experimental decay mode. If the half-lives are very close to each other, then more than one options are tabulated for decay, however, the lowest half-life among these entries is mentioned in the second last column. In the last column, experimental half-lives are given with their errors \cite{nndc}. \par

As far as the deviation from the experimental data is concerned it is indeed found of statistical nature. To demonstrate this kind of statistical variation, we calculate uncertainty using the following formula in our calculated Q values and correspondingly in Log T$_{1/2}$.

\begin{equation}\label{uncertainty}
  u = \sqrt{\frac{\sum(x_i - \mu)^2}{n(n-1)}}
\end{equation}
Where, $x_i$ is the $i^{th}$ reading in the data set, $\mu$ ia the mean of data set and n is the number of readings in the data set. The uncertainty in theoretical values are mentioned in Table 4 along with average uncertainty of concerned nuclei from experiment \cite{nndc}. The Table 4 evidently describes the veracity of our theoretical data and validate the accuracy of our prediction.

\begin{table}[!htbp]
 \centering
 \caption{Uncertainty in the Q values (MeV) and Log T$_{1/2}$ (s).}
 \resizebox{0.45\textwidth}{!}{%
 \begin{tabular}{l|c|c}
 \hline
 Quantity & Theoretical & Experimental \\
 \hline
Q$_{\alpha}$&      0.037  & 0.14  \\
Q$_{\beta^{-}}$&   0.038  & 0.37   \\
Q$_{\beta^{+}/EC}$&0.041  & 0.33   \\
\hline
Log T$_{1/2}$     & 0.22  & 0.23  \\
\hline
\end{tabular}}
\end{table}

Further, it is gratifying to note from Table 3 that RMF theory along with the applied phenomenological formulas and considering all the decay modes on equal footing, successfully able to reproduced experimental decay modes and half-lives for most of the considered nuclei. As can be seen from table that $\alpha$-decay and spontaneous fission come forth as good competitors and share most of the nuclei, however, the probability of weak-decay in these transfermium isotopes can not be disregarded. Our theoretical prediction indicates an equal probability of weak-decay as that of $\alpha$-decay and spontaneous fission in several nuclei as mentioned in Table 3. Therefore, our results affirm weak-decay in transfermium isotopes inline with the refs. \cite{hofmann2016,karpov2012,zagrebaev2012,heenen2015,hirsch1993,moller2019,sarriguren2019}.\par

\begin{figure}[h]
\centering
\includegraphics[width=0.6\textwidth]{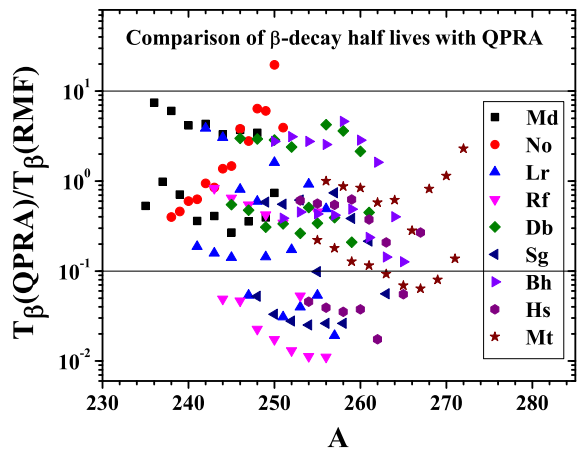}
\caption{(Colour online) Comparison of calculated $\beta$-decay half-lives with quasi-particle random-phase approximation (QRPA) results \cite{moller2019,moller1990}}.\label{fig2}
\end{figure}

Since our calculation of half-life for weak-decay is very crude and does not involve the main ingredients of weak-decay i.e. phase space factor and nuclear structure which is used to render energy distribution of the GT strength \cite{sarriguren2019}, therefore, testing of half-lives of weak-decay becomes essential for further prediction. For this trial, we compare our calculated half-lives of weak-decays with the recent calculations of Gamow–Teller $\beta$-decay rates which are obtained from a quasi-particle random-phase approximation with single-particle levels and wave functions at the calculated nuclear ground-state shapes as input quantities \cite{moller2019}. In Fig. \ref{fig2}, we show a comparison between half-lives calculated in this paper by using an empirical formula of Fiset and Nix \cite{Fiset1972} and the ones obtained from a quasi-particle random-phase approximation (QRPA) \cite{moller2019,moller1990}, and satisfactorily the comparison stands fairly reasonable.\par
The extensive comparison of our results in Tables 3 \& 4 along with Fig. \ref{fig2}, allows us to apply the whole formalism to the other isotopes of transfermium nuclei, by which prediction can be made for the most probable decay mode and correspondingly the value of half-life. In Table 5, we have tabulated Log of half-lives for $\alpha$, $\beta^\pm$, EC, and SF-decay along with probable decay mode(s) with its half-life. From a close watch, it is noticeable that $\alpha$-decay dominates this part of periodic chart, and the most of the half-lives of $\alpha$-decay are in current experimental reach. In addition, weak-decay is also found more likely in some of the nuclei apparently, and may provide useful inputs for future experiments on the road of search of new elements and weak-decays in this superheavy region.\par

\begin{center}
\small
\setlength{\tabcolsep}{3pt}
\begin{longtable}{|c|c|c|c|c|c|c|c|}
\caption{Prediction for most probable decay mode and calculated half-life for transfermium isotopes}\\
\hline
 \multicolumn{1}{|c}{Nucleus}&
 \multicolumn{1}{|c}{Log T$_\alpha$}&
  \multicolumn{1}{|c}{Log T$_{\beta^-}$}&
   \multicolumn{1}{|c}{Log T$_{\beta^+}$}&
  \multicolumn{1}{|c}{Log T$_{EC}$}&
   \multicolumn{1}{|c}{Log T$_{SF}$}&
 \multicolumn{1}{|c}{Decay-Mode}&
 \multicolumn{1}{|c|}{T$_{1/2}$}\\
\hline
\endfirsthead
\multicolumn{8}{c}%
{\tablename\ \thetable\ -- \textit{Continued from previous page}} \\
\hline
 \multicolumn{1}{|c}{Nucleus}&
 \multicolumn{1}{|c}{Log T$_\alpha$}&
  \multicolumn{1}{|c}{Log T$_{\beta^-}$}&
   \multicolumn{1}{|c}{Log T$_{\beta^+}$}&
  \multicolumn{1}{|c}{Log T$_{EC}$}&
   \multicolumn{1}{|c}{Log T$_{SF}$}&
 \multicolumn{1}{|c}{Decay-Mode}&
 \multicolumn{1}{|c|}{T$_{1/2}$}\\
\hline
\endhead
\hline \multicolumn{8}{r}{\textit{Continued on next page}} \\
\endfoot
\hline
\endlastfoot
$^{235}$Md  & -3.32   &  - &   0.37   & 1.85      &   -10.23  &   SF                          &   <10$^{-2}$   \\
$^{236}$Md  & -1.77   &  - &   -0.46  & 1.34      &   -10.30  &   SF                          &   <10$^{-2}$   \\
$^{237}$Md  & -1.36   &  - &   0.48   & 1.91      &   -8.46   &   SF                          &   <10$^{-2}$   \\
$^{238}$Md  & -0.97   &  - &   -0.37  & 1.38      &   -7.46   &   SF                          &   <10$^{-2}$   \\
$^{239}$Md  & -0.95   &  - &   0.68   & 2.00      &   -5.34   &   SF                          &   <10$^{-2}$   \\
$^{240}$Md  & -0.25   &  - &   -0.09  & 1.51      &   -4.59   &   SF                          &   <10$^{-2}$   \\
$^{241}$Md  & -0.32   &  - &   1.03   & 2.17      &   -2.36   &   SF                          &   <10$^{-2}$   \\
$^{242}$Md  & 0.24    &  - &   0.15   & 1.63      &   -1.38   &   SF                          &   0.04s      \\
$^{243}$Md  & 0.02    &  - &   1.29   & 2.29      &   0.82    &   $\alpha$                    &   1.06s      \\
$^{244}$Md  & 0.98    &  - &   0.52   & 1.80      &   1.48    &   $\beta^{+}/EC$/$\alpha$     &   3.29s      \\
$^{261}$Md  & 8.64    &  - &     -    &   -       &   8.61    &    SF/$\alpha$                &    12.88y    \\
$^{262}$Md  & 9.94    &  2.96&   -    &   -       &   8.49    &    $\beta^{-}$                &    15.14m    \\
$^{263}$Md  & 10.71   &  5.24&   -    &   -       &   9.89    &    $\beta^{-}$                &    48.30h    \\
$^{264}$Md  & 11.97   &  2.50&   -    &   -       &   8.75    &    $\beta^{-}$                &    5.25m     \\
$^{265}$Md  & 13.34   &  4.16&   -    &   -       &   8.83    &    $\beta^{-}$                &    4.02h     \\
$^{266}$Md  & 13.92   &  1.81&   -    &   -       &   7.52    &    $\beta^{-}$                &    1.07m     \\
$^{267}$Md  & 15.70   &  3.41&   -    &   -       &   7.52    &    $\beta^{-}$                &    43.20m    \\
$^{268}$Md  & 15.94   &  1.57&   -    &   -       &   6.35    &    $\beta^{-}$                &    37.43s    \\
\hline
$^{238}$No  &  -2.63   &  - &   0.74  &  0.82   &  -10.58  &   SF             &      <10$^{-2}$   \\
$^{239}$No  &  -1.60   &  - &   0.85  &  0.91   &  -7.92   &   SF             &      <10$^{-2}$   \\
$^{240}$No  &  -1.97   &  - &   0.71  &  0.79   &  -8.64   &   SF             &      <10$^{-2}$   \\
$^{241}$No  &  -1.63   &  - &   0.82  &  0.88   &  -6.05   &   SF             &      <10$^{-2}$   \\
$^{242}$No  &  -1.55   &  - &   0.63  &  0.73   &  -6.24   &   SF             &      <10$^{-2}$   \\
$^{243}$No  &  -1.63   &  - &   0.77  &  0.84   &  -3.21   &   SF             &      <10$^{-2}$   \\
$^{244}$No  &  -1.56   &  - &   0.58  &  0.68   &  -3.37   &   SF             &      <10$^{-2}$   \\
$^{245}$No  &  -0.50   &  - &   0.69  &  0.77   &  -0.37   &   $\alpha$       &       0.31s       \\
$^{246}$No  &  -0.81   &  - &   0.45  &  0.58   &  -0.68   &   $\alpha$       &       0.16s       \\
$^{247}$No  &  -0.56   &  - &   0.61  &  0.71   &  2.29    &   $\alpha$       &       0.27s       \\
$^{248}$No  &  -0.21   &  - &   0.32  &  0.49   &  1.87    &   $\alpha$       &       0.61s       \\
$^{249}$No  &  0.26    &  - &   0.52  &  0.64   &  4.90    &   $\alpha$       &       1.80s       \\
$^{261}$No &   6.97    &   - &    - &    -1.53  &  7.60    &   $\beta^{+}/EC$ &      0.03s        \\
$^{263}$No &   8.63    &  6.63&   - &      -     &  8.34    &   $\beta^{-}$    &      48.83d       \\
$^{264}$No &   8.55    &  - &    - &      -     &  7.52    &   SF             &      1.06y        \\
$^{265}$No &   11.58   &  5.10&   - &      -     &  9.59    &   $\beta^{-}$    &      1.45d        \\
$^{266}$No &   11.18   &  - &     - &     -     &  7.06    &   SF             &      0.99y        \\
$^{267}$No &   13.18   &  4.08 &   - &     -     &  7.97    &   $\beta^{-}$    &      3.33h        \\
$^{268}$No &   13.25   & - &      - &     -     &  4.94    &   SF             &      24.24h       \\
\hline
$^{241}$Lr  &   -3.62  &    -   &  0.28  &     1.74    &   -4.47   &    SF                    &  <10$^{-2}$     \\
$^{242}$Lr  &   -1.77  &    -   &  -0.49 &     1.25    &   -3.60   &    SF                    &  <10$^{-2}$          \\
$^{243}$Lr  &   -3.89  &    -   &  0.41  &     1.81    &   -6.69   &    SF                    &  <10$^{-2}$     \\
$^{244}$Lr  &   -1.96  &    -   &  -0.26 &     1.36    &   -6.01   &    SF                    &  <10$^{-2}$          \\
$^{245}$Lr  &   -3.25  &    -   &  0.65  &     1.92    &   -4.10   &    SF                    &   <10$^{-2}$    \\
$^{246}$Lr  &   -1.19  &     - &  -0.01 &     1.48    &   -3.36   &     SF                   &   <10$^{-2}$         \\
$^{247}$Lr  &   -1.37  &     - &  1.24  &     2.20    &   -1.45   &     SF                   &   <10$^{-2}$          \\
$^{248}$Lr  &   -0.77  &     - &  0.32  &     1.64    &   -0.96   &     SF                   &  <10$^{-2}$         \\
$^{249}$Lr  &   -0.70  &     - &  1.62  &     2.38    &   1.00    &     $\alpha$                   &   0.20s          \\
$^{250}$Lr  &   0.75   &     - &  0.69  &     1.81    &   1.97    &     $\beta^{+}/EC$/$\alpha$     &   4.86s          \\
$^{251}$Lr  &   0.38   &   -    &  2.06  &     2.58    &   3.90    &    $\alpha$                    &  2.39s          \\
$^{263}$Lr  &   5.34   &   -    &   -    &      -      &   7.28    &    $\alpha$                    &   60.51h        \\
$^{264}$Lr  &   6.25   &  4.45  &   -    &      4.33   &   7.74    &    $\beta^{+}/EC$/$\beta^{-}$   &   5.95h         \\
$^{265}$Lr  &   6.65   &   -    &   -    &      -      &   9.65    &    $\alpha$                    &   51.17d        \\
$^{267}$Lr  &   7.86   &  5.04  &   -    &      -      &   9.19    &    $\beta^{-}$                 &   30.25h        \\
$^{268}$Lr  &   8.17   &   -    &   -    &      -      &   7.40    &    SF                          &   0.79y         \\
$^{269}$Lr  &   8.89   &  3.77  &   -    &       -     &   7.06    &    $\beta^{-}$                 &   1.65h         \\
$^{270}$Lr  &   9.56   &  6.88  &   -    &       -     &   5.12    &    SF                          &   36.42h        \\
\hline
$^{243}$Rf &   -3.63  &   -     &  -0.07 &   1.55   &  -9.93   &   SF            &  <10$^{-2}$       \\
$^{244}$Rf &   -3.45  &   -     &  1.17  &   2.40   &  -10.48  &   SF            &  <10$^{-2}$       \\
$^{245}$Rf &   -2.95  &   -     &  0.13  &   1.64   &  -7.97   &   SF            &  <10$^{-2}$       \\
$^{246}$Rf &   -3.00  &   -     &  1.47  &   2.54   &  -8.07   &   SF            &  <10$^{-2}$       \\
$^{247}$Rf &   -1.47  &   -     &  0.37  &   1.76   &  -5.16   &   SF            &  <10$^{-2}$       \\
$^{248}$Rf &   -1.93  &   -     &  1.87  &   2.73   &  -5.54   &   SF            &  <10$^{-2}$       \\
$^{249}$Rf &   -1.69  &   -     &  0.65  &   1.90   &  -2.83   &   SF            &  <10$^{-2}$       \\
$^{250}$Rf &   -1.56  &   -     &  2.21  &   2.89   &  -2.86   &   SF            &  <10$^{-2}$       \\
$^{251}$Rf &   -0.58  &   -     &  1.02  &   2.07   &  0.16    &   $\alpha$      &  1.45s            \\
$^{252}$Rf &   -0.73  &   -     &  2.69  &   3.11   &  -0.10   &   $\alpha$      &  0.80s             \\
$^{264}$Rf &   3.87   &  4.14   &   -    &    -     &  5.47    &   $\alpha$      &  2.05h            \\
$^{266}$Rf &   5.23   &  4.66   &   -    &    -     &  8.27    &   $\beta^{-}$   &  12.63h           \\
$^{267}$Rf &   6.68   &   -     &   -    &    -     &  10.14   &   $\alpha$      &  55.17d           \\
$^{268}$Rf &   6.55   &   -     &   -    &    -     &  7.58    &   $\alpha$      &  40.77d            \\
$^{269}$Rf &   7.46   &  6.65   &   -    &    -     &  7.98    &   $\beta^{-}$   &  51.37d           \\
$^{270}$Rf &   6.95   &   -     &   -    &    -     &  5.06    &   SF            &  31.73h           \\
$^{271}$Rf &   7.65   &  4.69   &   -    &    -     &  5.62    &   $\beta^{-}$   &  13.46h           \\
$^{272}$Rf &   7.87   &   -     &   -    &    -     &  2.89    &   SF            &  13.08m           \\
\hline
$^{245}$Db  &  -4.76 &   -    &  0.05    &   1.56  &   -10.57 &    SF            &  <10$^{-2}$     \\
$^{246}$Db  &  -3.11 &   -    &  -0.68   &   1.08  &   -10.43 &    SF            &  <10$^{-2}$     \\
$^{247}$Db  &  -3.37 &   -    &  0.17    &   1.62  &   -8.29  &    SF            &  <10$^{-2}$     \\
$^{248}$Db  &  -2.11 &   -    &  -0.45   &   1.20  &   -7.97  &    SF            &  <10$^{-2}$     \\
$^{249}$Db  &  -2.65 &   -    &  0.53    &   1.79  &   -6.06  &    SF            &  <10$^{-2}$     \\
$^{250}$Db  &  -1.88 &   -    &  -0.24   &   1.30  &   -5.62  &    SF            &  <10$^{-2}$     \\
$^{251}$Db  &  -3.03 &   -    &  0.80    &   1.92  &   -3.22  &    SF            &  <10$^{-2}$     \\
$^{252}$Db  &  -1.29 &   -    &  0.04    &   1.44  &   -2.42  &    SF            &  <10$^{-2}$     \\
$^{253}$Db  &  -2.09 &   -    &  1.15    &   2.09  &   -0.32  &    $\alpha$      &  0.48s          \\
$^{254}$Db  &  -0.75 &   -    &  0.38    &   1.60  &   0.58   &    $\alpha$      &  3.77s           \\
$^{264}$Db  &  3.94  &   -    &  3.40    &   2.97  &   5.75   &    $\beta^{+}/EC$&   15.53m        \\
$^{265}$Db  &  3.32  &   -    &  8.46    &   4.49  &   7.69   &    $\alpha$      &   35.06m        \\
$^{266}$Db  &  4.27  &   -    &  3.92    &   3.19  &   8.71   &    $\beta^{+}/EC$&   25.78m        \\
$^{269}$Db  &  5.70  &   -    &   -      &    -    &   9.66   &    $\alpha$      &   5.84d          \\
$^{271}$Db  &  6.36  &   -    &   -      &    -    &   7.02   &    $\alpha$      &   26.54d        \\
$^{272}$Db  &  6.36  & 3.52   &  -       &   4.86  &   5.11   &    $\beta^{-}$   &   54.88m        \\
\hline
$^{248}$Sg  &  -3.89  &  -    &    0.94  &    2.21  &     -11.88  &   SF         &  <10$^{-2}$    \\
$^{249}$Sg  &  -2.54  &  -    &    0.00  &    1.50  &     -9.38   &   SF         &  <10$^{-2}$    \\
$^{250}$Sg  &  -3.14  &  -    &    1.32  &    2.39  &     -9.89   &   SF         &  <10$^{-2}$    \\
$^{251}$Sg  &  -2.97  &  -    &    0.23  &    1.61  &     -6.71   &   SF         &  <10$^{-2}$    \\
$^{252}$Sg  &  -2.93  &  -    &    1.61  &    2.53  &     -6.77   &   SF         &  <10$^{-2}$    \\
$^{253}$Sg  &  -2.05  &  -    &    0.50  &    1.74  &     -3.74   &   SF         &  <10$^{-2}$    \\
$^{254}$Sg  &  -2.16  &  -    &    1.98  &    2.70  &     -3.81   &   SF         &  <10$^{-2}$    \\
$^{255}$Sg  &  -1.33  &  -    &    0.85  &    1.91  &     -0.78   &   $\alpha$   &  0.05s         \\
$^{256}$Sg  &  -1.51  &  -    &    2.39  &    2.89  &     -0.75   &   $\alpha$   &  0.03s         \\
$^{257}$Sg  &  -0.65  &  -    &    1.19  &    2.07  &     2.25    &   $\alpha$   &  0.22s          \\
$^{267}$Sg  &  3.11   &  -    &    5.00  &    3.73  &     9.18    &   $\alpha$   &   21.24m       \\
$^{268}$Sg  &  3.29   &  -    &    -      &   -       &   8.61    &   $\alpha$   &   32.78m       \\
$^{270}$Sg  &  4.30   &  -    &    -      &   -       &   7.98    &   $\alpha$   &  5.58h         \\
$^{272}$Sg  &  4.60   &  -    &    -      &   -       &   5.31    &   $\alpha$   &  11.09h         \\
$^{273}$Sg  &  4.91   &  -    &    -      &   -       &   5.61    &   $\alpha$   &  22.63h        \\
$^{274}$Sg  &  5.05   &  -    &    -      &   -       &   2.62    &   SF         &  6.87m         \\
$^{275}$Sg  &  5.19   &  5.27 &    -      &   -       &   3.57    &   SF         &  61.37m        \\
$^{276}$Sg  &  5.13   &   -   &    -      &   -       &   1.27    &   SF         &  18.54s        \\
\hline
$^{250}$Bh  &   -2.80 &   -     &   -0.81  &  0.95  &  -12.09  &   SF          &  <10$^{-2}$              \\
$^{251}$Bh  &   -3.99 &   -     &   0.07   &  1.50  &  -9.62   &   SF          &  <10$^{-2}$              \\
$^{252}$Bh  &   -2.93 &   -     &   -0.63  &  1.04  &  -8.93   &   SF          &  <10$^{-2}$              \\
$^{253}$Bh  &   -3.73 &   -     &   0.30   &  1.61  &  -6.87   &   SF          &  <10$^{-2}$              \\
$^{254}$Bh  &   -2.04 &   -     &   -0.40  &  1.15  &  -6.03   &   SF          &  <10$^{-2}$               \\
$^{255}$Bh  &   -2.80 &   -     &   0.58   &  1.75  &  -4.04   &   SF          &  <10$^{-2}$              \\
$^{256}$Bh  &   -1.41 &   -     &   -0.12  &  1.28  &  -3.26   &   SF          &  <10$^{-2}$              \\
$^{257}$Bh  &   -2.26 &   -     &   0.88   &  1.89  &  -0.94   &   $\alpha$    &  0.01s         \\
$^{258}$Bh  &   -0.62 &   -     &   0.16   &  1.42  &  -0.22   &   $\alpha$    &  0.24s          \\
$^{259}$Bh  &   -0.83 &   -     &   1.25   &  2.06  &  1.49    &   $\alpha$    &  0.15s         \\
$^{273}$Bh  &   4.02  &   -     &    -     &    -   &  7.15    &   $\alpha$    &  2.92h         \\
$^{275}$Bh  &   4.25  &   -     &    -     &    -   &  4.15    &   SF/$\alpha$ &  3.88h         \\
$^{276}$Bh  &   4.83  &   5.09 &    7.12   &   4.14 &  3.12    &   SF          &  21.91m        \\
$^{277}$Bh  &   4.65  &   -    &     -     &   -    &  3.02    &   SF          &  17.52m        \\
$^{278}$Bh  &   5.23  &  4.07 &      -     &  4.97  &  2.63    &   SF          &  7.05m         \\
\hline
$^{253}$Hs  &  -3.74  &  -  &  -0.17 &     1.34    &  -10.25  &  SF         & <10$^{-2}$   \\
$^{254}$Hs  &  -3.87  &  -  &  1.07  &     2.20    &  -10.38  &  SF         & <10$^{-2}$   \\
$^{255}$Hs  &  -2.66  &  -  &  0.09  &     1.47    &  -7.50   &  SF         & <10$^{-2}$   \\
$^{256}$Hs  &  -3.16  &  -  &  1.35  &     2.33    &  -7.64   &  SF         & <10$^{-2}$   \\
$^{257}$Hs  &  -2.39  &  -  &  0.32  &     1.58    &  -4.83   &  SF         & <10$^{-2}$   \\
$^{258}$Hs  &  -2.56  &  -  &  1.63  &     2.47    &  -4.61   &  SF         & <10$^{-2}$   \\
$^{259}$Hs  &  -1.73  &  -  &  0.53  &     1.68    &  -1.72   &  SF/$\alpha$ &  0.02s       \\
$^{260}$Hs  &  -1.45  &  -  &  1.96  &     2.62    &  -1.68   &  SF         &  0.02s       \\
$^{261}$Hs  &  -0.28  &  -  &  0.99  &     1.90    &  0.85    &  $\alpha$   &  0.53s       \\
$^{262}$Hs  &  -1.06  &  -  &  2.47  &     2.86    &  0.22    &  $\alpha$   &  0.09s        \\
$^{271}$Hs  &  2.56   &  -  &  3.83  &     3.19    &  8.86    &  $\alpha$   &  5.99m       \\
$^{272}$Hs  &  2.00   &  -  &   -      &     5.31  &  6.98    &  $\alpha$   &  1.68m       \\
$^{274}$Hs  &  1.81   &  -  &   -      &    -      &  4.77    &  $\alpha$   &  63.86s      \\
$^{276}$Hs  &  1.93   &  -  &   -      &    -      &  2.28    &  $\alpha$   &  85.50s       \\
$^{278}$Hs  &  2.06   &  -  &   -      &    -      &  2.71    &  $\alpha$   &  1.92m       \\
$^{279}$Hs  &  3.61   &  -  &   -      &     5.66  &  3.45    &  SF/$\alpha$ &  47.37m      \\
$^{280}$Hs  &  2.96   &  -  &   -      &    -      &  2.47    &  SF/$\alpha$ &  4.87m        \\
$^{281}$Hs  &  4.18   & -   &   -      &    -      &  5.42    &  $\alpha$   &  4.19h       \\
$^{282}$Hs  &  4.45   & -   &   -      &    -      &  3.79    &  SF         &  1.70h       \\
\hline
$^{255}$Mt &  -4.31  &  -     &    -0.10  &    1.34    &  -10.76   &   SF                        &  <10$^{-2}$      \\
$^{256}$Mt &  -2.35  &  -     &    -0.69  &    0.93    &  -10.21   &   SF                        &  <10$^{-2}$      \\
$^{257}$Mt &  -3.82  &  -     &    0.14   &    1.46    &  -7.99    &   SF                        &  <10$^{-2}$      \\
$^{258}$Mt &  -2.21  &  -     &    -0.52  &    1.01    &  -7.51    &   SF                        &  <10$^{-2}$      \\
$^{259}$Mt &  -3.10  &  -     &    0.41   &    1.59    &  -4.80    &   SF                        &  <10$^{-2}$      \\
$^{260}$Mt &  -1.65  &  -     &    -0.32  &    1.11    &  -4.03    &   SF                        &  <10$^{-2}$      \\
$^{261}$Mt &  -2.25  &  -     &    0.59   &    1.67    &  -1.70    &   $\alpha$                  &0.01s              \\
$^{262}$Mt &  -0.89  &  -     &    0.00   &    1.27    &  -1.20    &   SF                        &0.06s             \\
$^{263}$Mt &  -2.51  &  -     &    0.89   &    1.82    &  0.39     &   $\alpha$                  &  <10$^{-2}$      \\
$^{264}$Mt &  -1.58  &  -     &    0.21   &    1.37    &  0.50     &   $\alpha$                  & 0.03s             \\
$^{265}$Mt &  -2.67  &  -     &    1.28   &    2.00    &  1.35     &   $\alpha$                  &  <10$^{-2}$      \\
$^{269}$Mt &  -1.41  &  -     &    2.05   &    2.36    &  4.71     &   $\alpha$                  & 0.04s            \\
$^{271}$Mt &  -0.75  &  -     &    2.67   &    2.65    &  7.61     &   $\alpha$                  & 0.18s            \\
$^{272}$Mt &  0.35   &  -     &    1.79   &    2.11    &  7.71     &   $\alpha$                  & 2.25s             \\
$^{273}$Mt &  -0.16  &  -     &    3.68   &    3.09    &  7.85     &   $\alpha$                  & 0.69s            \\
$^{279}$Mt &  0.54   &  -     &      -    &    6.05    &   2.64    &   $\alpha$                  & 3.47s             \\
$^{280}$Mt &  1.04   &  6.32  &    3.88   &    3.04    &   2.65    &   $\alpha$                  & 11.03s             \\
$^{281}$Mt &  0.78   &   -     &    -     &      -     &   3.65    &   $\alpha$                  & 6.01s            \\
$^{282}$Mt &  3.62   &  6.61  &    6.10   &    3.87    &   4.17    &   $\alpha$                  &  1.16h           \\
$^{283}$Mt &  4.37   &   -     &   5.24   &      -     &   5.44    &   $\alpha$                  &  6.45h           \\
$^{284}$Mt &  6.58   &  4.95  &     -     &    4.94    &   6.34    &   $\beta^{+}/EC$/$\beta^{-}$&  24.36h          \\
\hline

\end{longtable}
\end{center}
\normalsize

\begin{figure}[h]
\centering
\includegraphics[width=1.0\textwidth]{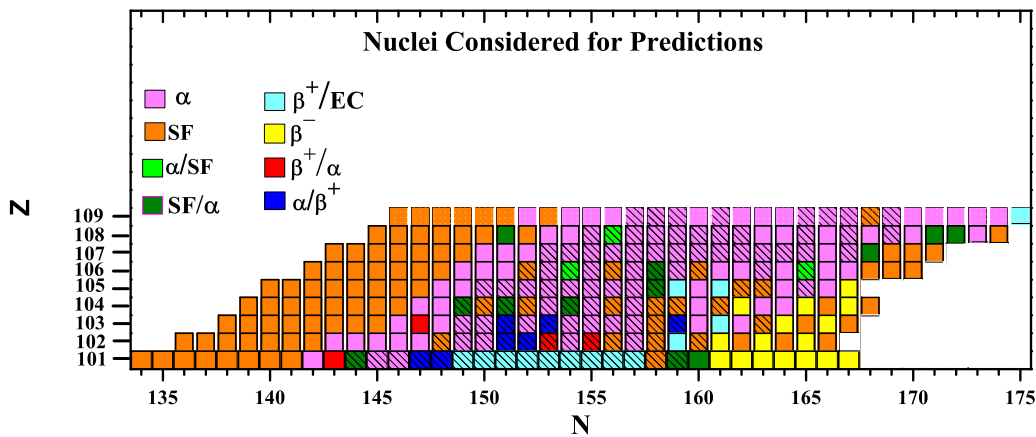}
\caption{(Colour online) Chart of considered nuclei with their probable decay modes.}.\label{fig3}
\end{figure}

All the prediction are summarized in form of a nuclear chart which is shown in Fig. \ref{fig3}. Dominant decay modes are shown by different colours for the considered nuclei. The shaded blocks correspond to decay modes which are known experimentally \cite{nndc}.

\section{Conclusions}
Decay modes are studied for odd and even transfermium nuclei in the range of $^{235-268}$Md, $^{238-268}$No, $^{241-270}$Lr, $^{243-272}$Rf, $^{245-272}$Db, $^{248-276}$Sg, $^{250-278}$Bh, $^{253-282}$Hs, and $^{255-284}$Mt. For all these nuclei, the calculations are done using relativistic mean-field theory (RMF) which are found in an excellent match with available experimental data and also with Hartree-Fock-Bogoliubov (HFB) mass model with HFB-24 functional, relativistic continuum Hartree-Bogoliubov (RCHB) theory with the relativistic density functional PC-PK1, nuclear mass table with the global mass formula WS4, and, recently reported Finite Range Droplet Model (FRDM) calculations. A comparison among $\alpha$, $\beta^\pm$, electronic capture decays, and spontaneous fission is demonstrated which leads to the most probable decay mode along with its half-life. As an important consequence, in spite of the fact that the $\alpha$-decay and SF modes are found to dominate, chances of weak-decay modes can not be ignored in considered transfermium nuclei. Indeed, we have found several isotopes in which weak-decay mode is quite comparable or sometimes more probable than $\alpha$-decay. The half-lives for weak-decay mode are found in accord with the half-lives calculated by using quasi-particle random-phase approximation (QRPA). These important findings suggest a new path to locate the gap between the nuclei synthesized by cold and hot reactions.

\section{Acknowledgement}
The authors take great pleasure in thanking the referee for his several suggestions and comments which helped to improve the manuscript. G. Saxena acknowledges the support provided by SERB (DST), Govt. of India under CRG/2019/001851.

\end{document}